\documentclass[prl,aps,amsfonts,amsmath,amssymb,nofootinbib,twocolumn,superscriptaddress]{revtex4}

\usepackage{graphicx}% Include figure files
\usepackage{bm}% bold math
\usepackage{hyperref}
\usepackage{IEEEtrantools}
\usepackage{natbib}
\usepackage{float}

\restylefloat{table}

\begin{document}

\title{Superconductivity mystery turns 25}

\author{N. Peter Armitage}\email{npa@jhu.edu}
\affiliation{Department of Physics and Astronomy, The Johns Hopkins University, Baltimore, Maryland 21218, USA}

\date{December 19, 2019}

\maketitle

\textbf{In 1994, an unconventional form of superconductivity was detected in strontium ruthenate. The discovery has shed light on the mechanism of unconventional superconductivity at high temperatures.  } 

\begin{center}  {\it  ``The great tragedy of Science [is] the slaying of a beautiful hypothesis by a [experimental] fact}."  \\ - T.H. Huxley,  Biogenesis and abiogenesis, (1884)

\end{center}

\bigskip

Superconductivity is an effect in which a material’s electrical resistance vanishes and any magnetic field is expelled below a transition temperature. Despite the remarkable phenomenology, this behaviour is actually quite common: almost half the elements in the periodic table are superconductors \cite{Shimizu11a}, albeit at temperatures near or below the extremely low one at which helium gas liquefies (about 4 kelvin). Since Nobel-prizewinning work in the late 1950s, we have had a successful theory \cite{Bardeen57a} of superconductivity in these conventional systems. Electrons bind into ‘Cooper pairs’ that have isotropic (direction-independent) properties through an interaction with vibrations of surrounding ions. Over the past 40 years, researchers have looked for unconventional superconductors that involve different pairing interactions, such as magnetic ones. In 1994, Maeno et al. \cite{Maeno94a} reported one of the clearest examples of unconventional superconductivity, in strontium ruthenate near 1 K.

Understanding unconventional superconductors requires identifying both the pairing interaction and the order parameter - a quantity that reflects the interaction and the macroscopic, typically anisotropic, properties of the unconventional superconductivity. The most substantial development in this area of study was the discovery of superconductivity in layered copper-oxide compounds (known as cuprates) in the mid-to-late 1980s. The phenomenon was detected \cite{Bednorz86a} at the unprecedentedly high temperature (for that time) of 30 K, which led to a worldwide effort to understand the mechanism of cuprate superconductivity.

The cuprates are now thought to have a highly anisotropic order parameter, and to have Cooper pairs made of electrons that have anti-aligned spins (intrinsic angular momenta). Such spins form non-magnetic states that have even parity, which means that the wavefunction of the state does not change sign if the signs of the spatial coordinates are flipped. Cuprate superconductivity has been proposed \cite{Nagaosa97a} to arise from an interaction of electrons with antiferromagnetic spin fluctuations (antiferromagnetism is a form of magnetism in which spins are anti-aligned with their neighbours). However, no theory has yet gained general acceptance.

One method that has been used to try to understand these compounds is to search for superconductivity in materials that are related in some way to the cuprates. In this way, it might be possible to identify the structural, electronic or magnetic features that are essential for the materials’ high transition temperatures. In particular, the cuprate discovery led to a huge effort to investigate compounds that contain transition metals other than copper.

\begin{figure*}
\includegraphics[width=12cm]{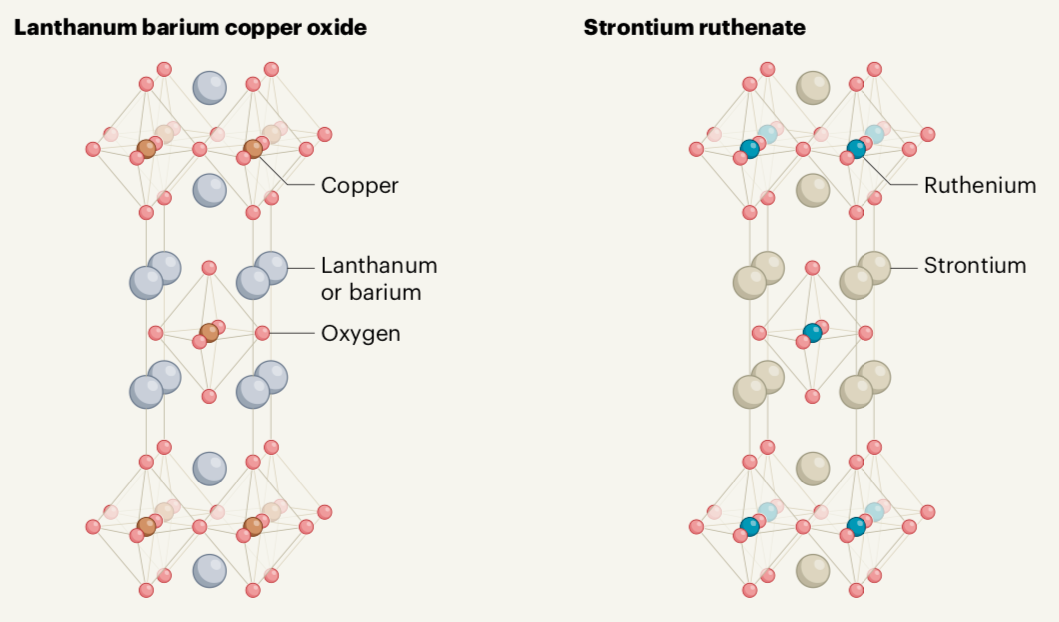}
\caption{\label{tempdependence} \textbf{Crystal structures of two superconductors.}   In 1986, lanthanum barium copper oxide was found \cite{Bednorz86a} to superconduct (transport electricity without resistance) at the relatively high temperature of 30 kelvin. Eight years later, Maeno et al. \cite{Maeno94a} reported the discovery of superconductivity in strontium ruthenate at about 1 K. Although these two materials have the same crystal structures at high temperatures, their superconductivity mechanisms are likely to be markedly different.}
\end{figure*}

It was against this backdrop that Maeno and colleagues found superconductivity in strontium ruthenate, at about 1 K. This was decidedly not high-temperature superconductivity. But the work caused tremendous excitement because it described the detection of superconductivity in another layered transition-metal oxide — and in a material that has the same crystal structure as the original superconducting cuprate, lanthanum barium copper oxide \cite{Maeno94a} (Fig. 1). Almost immediately, it was realized that there were both similarities and differences between the cuprates and strontium ruthenate.

One main difference is that pure compounds of the cuprates (such as lanthanum copper oxide) are antiferromagnetic insulators and require the substitution of atoms (such as barium for lanthanum) to conduct electricity. By contrast, pure strontium ruthenate is strongly metallic. A striking aspect of the superconducting cuprates is that their metallic state at temperatures above the transition temperature seems to be even more unconventional than their superconducting state. The metallic state is thought to be the result of strong interactions between electrons. A radically new theory of ‘strange metals’ might be needed to understand the high-temperature metallic state and thereby also the superconducting state that forms from it \cite{Hussey08a}. In strontium ruthenate, electron interactions are also strong, but they do not change the fundamental character of the metallic state.

This aspect, and the fact that related materials in the larger ruthenate family exhibit ferromagnetism (a form of magnetism in which spins are aligned with their neighbours), led to the proposal \cite{Rice95a} in 1995 that superconductivity in strontium ruthenate could be an analogue of the superfluid A phase in helium-3. In this phase, the compound exists as a superfluid (a zero-viscosity liquid) made from odd-parity Cooper pairs of neutral helium-3 atoms that have aligned spins \cite{Lee97a}. The proposal gained much support, both for the compelling science that suggests it and for the beautiful idea that there could be an odd-parity superconductor driven by ferromagnetism in the same way that the cuprates might be even-parity superconductors driven by antiferromagnetism. Of course, the ``great tragedy of Science [is] the slaying of a beautiful hypothesis" by experimental facts \cite{Huxley84a}. Experiments always have the final say.

The exciting science, the ability to grow large, extremely pure crystals and an exceedingly collaborative research community pushed superconducting strontium ruthenate forward as a highly active topic of investigation. Moreover, there was the abiding sense that it should be possible to unambiguously determine the nature of the material’s unconventional order parameter, because its high-temperature metallic state — unlike that of the cuprates — seemed to obey the conventional theory of metals. This determination is an ongoing saga, with field-changing results coming even this year. Notable early work showed evidence for unconventional odd-parity pairing of electrons in nuclear magnetic resonance (NMR) spectroscopy \cite{Ishida11a}, and for spontaneous generation of magnetism \cite{Luke98a,Xia06a} consistent with the proposal outlined above.

In the past five years, sophisticated measurements of strontium ruthenate have failed to show an odd-parity superconducting transition splitting into two under mechanical strain, as had been predicted \cite{Hicks14a}. These measurements, along with a reinvestigation using NMR spectroscopy \cite{Pustogow19a}, have given compelling evidence that the superconductivity is likely to be even parity. But this even-parity state is inconsistent with the experiments that showed the presence of spontaneous magnetism. Therefore, the nature of unconventional superconductivity in strontium ruthenate must be considered unresolved

This problem, together with that of the cuprates, has pushed theory, experiment and materials synthesis forward in directions that would have been unimaginable when superconductivity in these compounds was discovered. And as is so often the case, many of the ideas that scientists have grappled with in the context of a hard problem have turned out to be incredibly influential in areas well beyond their original scope. In this particular case, important cross-fertilizing connections can be made with topological insulators (bulk electrical insulators that have conducting surfaces) and quantum computation \cite{Sato17a}. The research community is still hard at work on the mystery of strontium ruthenate. Experiments always have the final say.

 \bibliography{Sr214}

\begin{thebibliography}{10}
\expandafter\ifx\csname url\endcsname\relax
  \def\url#1{\texttt{#1}}\fi
\expandafter\ifx\csname urlprefix\endcsname\relax\def\urlprefix{URL }\fi
\providecommand{\bibinfo}[2]{#2}
\providecommand{\eprint}[2][]{\url{#2}}

\bibitem{Shimizu11a}
\bibinfo{author}{Shimizu, K.}
\newblock \emph{\bibinfo{title}{100 Years of Superconductivity}}
  (\bibinfo{publisher}{Taylor \& Francis}, \bibinfo{year}{2011}).

\bibitem{Bardeen57a}
\bibinfo{author}{Bardeen, J.}, \bibinfo{author}{Cooper, L.~N.} \&
  \bibinfo{author}{Schrieffer, J.~R.}
\newblock \bibinfo{title}{Theory of superconductivity}.
\newblock \emph{\bibinfo{journal}{Physical review}}
  \textbf{\bibinfo{volume}{108}}, \bibinfo{pages}{1175} (\bibinfo{year}{1957}).

\bibitem{Maeno94a}
\bibinfo{author}{Maeno, Y.} \emph{et~al.}
\newblock \bibinfo{title}{Superconductivity in a layered perovskite without
  copper}.
\newblock \emph{\bibinfo{journal}{Nature}} \textbf{\bibinfo{volume}{372}},
  \bibinfo{pages}{532--534} (\bibinfo{year}{1994}).

\bibitem{Bednorz86a}
\bibinfo{author}{Bednorz, J.~G.} \& \bibinfo{author}{M{\"u}ller, K.~A.}
\newblock \bibinfo{title}{{Possible highT$_c$ superconductivity in the Ba- La-
  Cu- O system}}.
\newblock \emph{\bibinfo{journal}{Zeitschrift f{\"u}r Physik B Condensed
  Matter}} \textbf{\bibinfo{volume}{64}}, \bibinfo{pages}{189--193}
  (\bibinfo{year}{1986}).

\bibitem{Nagaosa97a}
\bibinfo{author}{Nagaosa, N.}
\newblock \bibinfo{title}{{Superconductivity and antiferromagnetism in
  high-T$_c$ cuprates}}.
\newblock \emph{\bibinfo{journal}{Science}} \textbf{\bibinfo{volume}{275}},
  \bibinfo{pages}{1078--1079} (\bibinfo{year}{1997}).

\bibitem{Hussey08a}
\bibinfo{author}{Hussey, N.~E.}
\newblock \bibinfo{title}{{Phenomenology of the normal state in-plane transport
  properties of high-T$_c$ cuprates}}.
\newblock \emph{\bibinfo{journal}{Journal of Physics: Condensed Matter}}
  \textbf{\bibinfo{volume}{20}}, \bibinfo{pages}{123201}
  (\bibinfo{year}{2008}).

\bibitem{Rice95a}
\bibinfo{author}{Rice, T.} \& \bibinfo{author}{Sigrist, M.}
\newblock \bibinfo{title}{{Sr$_2$RuO$_4$: an electronic analogue of 3He?}}
\newblock \emph{\bibinfo{journal}{Journal of Physics: Condensed Matter}}
  \textbf{\bibinfo{volume}{7}}, \bibinfo{pages}{L643} (\bibinfo{year}{1995}).

\bibitem{Lee97a}
\bibinfo{author}{Lee, D.~M.}
\newblock \bibinfo{title}{The extraordinary phases of liquid 3he}.
\newblock \emph{\bibinfo{journal}{Reviews of Modern Physics}}
  \textbf{\bibinfo{volume}{69}}, \bibinfo{pages}{645} (\bibinfo{year}{1997}).

\bibitem{Huxley84a}
\bibinfo{author}{Huxley, T.~H.}
\newblock \bibinfo{title}{{Biogenesis and abiogenesis; collected essays, Vol.
  8}} (\bibinfo{year}{1884}).

\bibitem{Ishida11a}
\bibinfo{author}{Ishida, K.} \emph{et~al.}
\newblock \bibinfo{title}{{Ru NMR probe of spin susceptibility in the
  superconducting state of Sr$_2$RuO$_4$}}.
\newblock \emph{\bibinfo{journal}{Physical Review B}}
  \textbf{\bibinfo{volume}{63}}, \bibinfo{pages}{060507}
  (\bibinfo{year}{2001}).

\bibitem{Luke98a}
\bibinfo{author}{Luke, G.~M.} \emph{et~al.}
\newblock \bibinfo{title}{{Time-reversal symmetry-breaking superconductivity in
  Sr$_2$RuO$_4$}}.
\newblock \emph{\bibinfo{journal}{Nature}} \textbf{\bibinfo{volume}{394}},
  \bibinfo{pages}{558--561} (\bibinfo{year}{1998}).

\bibitem{Xia06a}
\bibinfo{author}{Xia, J.}, \bibinfo{author}{Maeno, Y.},
  \bibinfo{author}{Beyersdorf, P.~T.}, \bibinfo{author}{Fejer, M.} \&
  \bibinfo{author}{Kapitulnik, A.}
\newblock \bibinfo{title}{{High resolution polar Kerr effect measurements of
  Sr$_2$RuO$_4$: Evidence for broken time-reversal symmetry in the
  superconducting state}}.
\newblock \emph{\bibinfo{journal}{Physical review letters}}
  \textbf{\bibinfo{volume}{97}}, \bibinfo{pages}{167002}
  (\bibinfo{year}{2006}).

\bibitem{Hicks14a}
\bibinfo{author}{Hicks, C.~W.} \emph{et~al.}
\newblock \bibinfo{title}{{Strong increase of T$_c$ of Sr$_2$RuO$_4$ under both
  tensile and compressive strain}}.
\newblock \emph{\bibinfo{journal}{Science}} \textbf{\bibinfo{volume}{344}},
  \bibinfo{pages}{283--285} (\bibinfo{year}{2014}).

\bibitem{Pustogow19a}
\bibinfo{author}{Pustogow, A.} \emph{et~al.}
\newblock \bibinfo{title}{Constraints on the superconducting order parameter in
  sr$_2$ruo$_4$ from oxygen-17 nuclear magnetic resonance}.
\newblock \emph{\bibinfo{journal}{Nature}} \textbf{\bibinfo{volume}{574}},
  \bibinfo{pages}{72--75} (\bibinfo{year}{2019}).

\bibitem{Sato17a}
\bibinfo{author}{Sato, M.} \& \bibinfo{author}{Ando, Y.}
\newblock \bibinfo{title}{Topological superconductors: a review}.
\newblock \emph{\bibinfo{journal}{Reports on Progress in Physics}}
  \textbf{\bibinfo{volume}{80}}, \bibinfo{pages}{076501}
  (\bibinfo{year}{2017}).

\end{thebibliography}

\normalsize
 
\end{document}